# A Formal Approach to Distributed System Security Test Generation

Vladimir A. Khlevnoy, Andrey A. Shchurov,

*Department of Cybernetics, Faculty of Electrical Engineering,*
*Czech Technical University in Prague, The Czech Republic*

*Abstract*— **Deployment of distributed systems sets high requirements for procedures for the security testing of these systems. This work introduces: (1) a list of typical threats based on standards and actual practices; (2) an extended six-layered model for test generation mission on the basis of technical specifications and end-user requirements. Based on the list of typical threats and the multilayer model, we describe a formal approach to the automated design and generation of security mechanisms checklists for complex distributed systems.**

## I. INTRODUCTION

*If you can't describe what you are doing as a process, you don't know what you're doing.*

— *William Edwards Deming*

Nowadays distributed systems have critical security requirements. Their failure may endanger human lives and the environment, do serious damage to major economic infrastructure, endanger personal privacy, undermine the viability of whole business sectors and facilitate crime [1]. As a consequence, the most difficult part of distributed systems deployment is the question of assurance (whether the system will work) and verification. If assurance is difficult, verification is even more difficult: it is a question of how to convince customers (and, in extremis, a jury) that a system is indeed fit for its goals including security objectives.

Generally, security requirements differ greatly from one system to another. But in the real world many systems have failed because their designers: (1) had protected the wrong things; (2) had protected the right things but in the wrong way; (3) or some things had been just simply forgotten. As a possible solution, it is necessary to determine a formal list of control objectives during the design phase of the System Development Life Cycle (SDLC) and, as the next step, to show that each component of this list meets at least one protection mechanism during the implementation phase of the SDLC: i.e. it is necessary to have checklists.

But we face the great challenge of the dual nature of distributed systems. In fact, it is necessary to pay respect to their software nature – services and end-user application. On the other hand, the distribution character of these systems forces us to consider their network nature. Historically, we have two independent administrative domains with two different approaches to security policies: (1) the communication (network-based) domain [2]; and (2) the system (software-based) domain [3] instead of comprehensive (integral) approach.

Our main goal is the automated design and generation of security mechanisms checklists (or a set of test cases) for distributed systems based on end-user requirements and technical specifications as a necessary part of project documentation. We need to state here - working engineers treat formal methods as they are widely taught in universities and not used anywhere in the real world. But in the case of complex or non-standard systems, personal experience and/or intuition are often inadequate. Thus, to accomplish such a goal we need to identify a formal list of control objectives with the following criteria: (1) it should be based on standards and/or well-known practical methods (as a formal document); (2) it has to cover all aspects of distributed systems; and (3) it has to be simple enough for practical application.

The rest of this paper is structured as follows. Section 2 introduces the related work. Section 3 presents the extended six-layered model of distributed systems for test generation mission and the checklists generation approach. Section 4 introduces an example based on a simple information system. Finally, conclusion remarks are given in Section 5.

## II. RELATED WORK

The current revision of ITU-T X.805 [4] standard defines the security architecture that addresses three essential questions with regard to end-to-end security:

- What kind of protection is needed and against what threats?
- What are the distinct types of system equipment and facility groupings that need to be protected?
- What are the distinct types of system activities that need to be protected?

### A. Definition of typical threats

In respect to our main goals, the answer on the first question must be based on related standards or well-known practical methods.

The current revision of ISO/IEC 27005:2011 [5] standard contains a list of typical threats and can be used as a starting point. We need to state here – this list covers both aspects (software-based and network-based) of distributed systems but not only these aspects.

On the other hand as recent practical approaches, Trilateral Research & Consulting describes more than 30 different risk management standards and methodologies [6]. The most widespread and comprehensive solutions are [7]: EBIOS [8],

IRAM [9], IT-Grundschutz [10], MAGERIT [11], Mehari [12] and OCTAVE [13].

### B. Definition of protected objects

In respect to our main goals, the answer to the later questions must be based on system specifications. The search for necessary system equipment and system activities usually involves analyzing system models, with the analysis covering paths in a model. In this context, this work lies in the area of model-based testing (MBT) based on formal specifications.

Bernot et al [14] set up a theoretical basis for specification-based testing, explaining how a formal specification can serve as a base for test case generation. Dick and Faivre [15] propose transforming formal specifications into a disjunctive normal form (DNF) and then using it as the basis for test case generation. Donat [16] represents a technique for automatic transformation of formal specifications into test templates and taxonomy for coverage schemes. Hong et al [17] show how coverage criteria based on control-flow or data-flow properties can be specified as sets of temporal logic formulas, including state and transition coverage as well as criteria based on definition-use pairs. Liu and Shen [18] describe a method that can be used for (1) identifying all interface scenarios, formalizing requirements into formal operation specifications whose interfaces are consistent with the corresponding ones of the program; and (2) for testing programs based upon the formal specifications (scenario-coverage strategy). In turn, Shchurov and Marik [19] present a requirements-coverage test strategy that covers both hardware-based (system equipment) and software-based (system activities) aspects of complex distributed systems.

### III. CHECKLISTS GENERATION

### A. Basic approach

The essential idea of our approach is based on:

- IT-Grundschutz risk management methodology [10];
- component-based approach with its two important consequences: (1) components are built to be reused in different systems, and (2) component development process is separated from the system development process [20], [21].

In our case IT-Grundschutz was chosen as the basic analytic tool based on the following notions:

- the list of typical threats of IT-Grundschutz is compatible with the list of typical threats of ISO/IEC 27005:2011;
- in contrast with ISO/IEC 27005:2011 and other solutions, the list of typical threats of IT-Grundschutz has a linear structure (the list of ISO/IEC 27005:2011 has a tree structure);
- in contrast with ISO/IEC 27005:2011, the list of typical threats of IT-Grundschutz has detailed descriptions for each threat.

We need to state here – the international standard ISO/IEC 27005:2011 and/or regional standards/methodologies can be used as an analytic tool depending on the state legislation and/or corporate requirements.

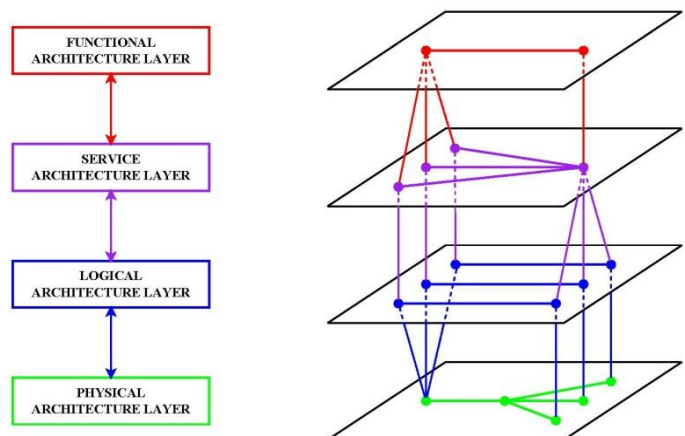

Fig. 1  Four-layered model of distributed systems for test generation missions [22]

In turn, the component-based approach refers to the fact that the functional usefulness of distributed systems does not depend on any particular part of these systems, but emerges from the way in which their components interact. Thus, a formal four layered model for test generation missions [22] can be used as a starting point. This model is stated as a four-layered graph as follows see Fig.1:

- The ready-for-use system architecture layer defines functional components and their interconnections.
- The service architecture layer defines software-based components (services/applications) and their interconnections.
- The logical architecture layer defines logical (virtual) components and their interconnections.
- The physical architecture layer defines hardware (physical) components and their interconnections.
- The interlayer projections define all types of components hierarchical (interlayer) relations/mapping. These relations make the layered model consistent and represent interlayer technologies (virtualization, clustering, etc.) used to build distributed systems.

Unfortunately, this multilayer model of complex systems includes only four layers and, as a consequence, does not cover the lists of typical threats completely. This problem can be solved by two additional layers:

- The engineering environment architecture layer. This layer defines external engineering systems (power supply systems, climate control systems, physical security systems, etc.) that are vital for normal operation of distributed systems and their interconnections. It is based on topological models (TMs) [23], where all systems (engineering and distributed) are represented as individual components.
- The social environment architecture layer. This layer defines an enterprise's organization infrastructures or "human networks". It is also based on TMs but represents persons or groups of persons and their working relationships.

These additional layers lie beyond the ISO/OSI Reference Model (RM) [24] but they provide a necessary complement to it with regard to our main goals. The final model is shown in Fig.2.

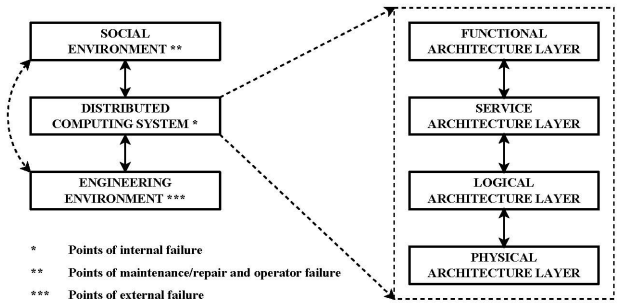

Fig. 2 Extended six-layered model of distributed systems

### B. Formal notations

Formal verification offers a rich toolbox of mathematical techniques that can support the model-based testing of computer systems. This toolbox contains logic programming as one of the most relevant techniques of model checking [25]. In turn, logic programming deals with logical facts and, as a consequence, the basic step is to determine the formal notations which make the layered model applicable for logic programming.

Applying of the requirements-coverage test strategy [19] to the system model (see Fig.2.) provides the set of system objects that need to be protected.

**Definition 1 (objects):** *Let the set $O_n$ denote the system objects for each layer n that need to be protected:*

$$O_n = V_n \cup T_n$$

where $V_n$ is a subset of individual components (or system equipment and facility groupings) on layer n; and $T_n$ is a subset of data flows (or system activities) between pairs of individual components on layer n, which must communicate.

As the next step, we have to partition the set of typical threats with regard to the model layered structure and the objects definition (see Definition 1).

**Definition 2 (threats/dangers):** *Let the set $Th_n$ denote the threats for each layer n:*

$$Th_n = Th_{n1} \cup Th_{n2}$$

where $Th_{n1}$ is a subset of threats that affects individual components on layer n; and $Th_{n2}$ is a subset of threats that affects data flows on layer n. In turn:

$$Th = \bigcup_{n=0}^{N} Th_n$$

where Th is the list of typical threats; and N is the number of layers of the system model.

In the case of IT-Grundschutz risk management methodology [10] sets of threats shown in the Table 1.

As the final step, we have to determine the security checklist generation process.

**Definition 3 (the security checklist):** *The set of test cases (or security checklist), denoted by $S_n$, is the mapping from the set of threats (see Definition 2) to the set of protected objects (see Definition 1) for every layer n:*

$$S_n : Th_n \to O_n$$

*In turn:*

$$S_n = S_{n1} \cup S_{n2}$$
$$S_{n1} : Th_{n1} \to V_n$$
$$S_{n2} : Th_{n2} \to T_n$$

where $S_{n1}$ is a subset of test cases that affects individual components on layer n; and $S_{n2}$ is a subset of test cases that affects data flows between pairs of individual components on layer n, which must communicate.

Generally, each individual component of $Th_n$ must be mapped to at least one component of $O_n$. As a consequence, $S_n$ cannot be an empty set without any exception. Then:

$$S = \bigcup_{n=0}^{N} S_n$$

where S is the completed set of test cases (or final security checklist); and N is the number of layers of the system model.

The graphical representation of the security checklist generation process is shown in Fig. 3.

### C. Complexity of the approach

The total number of test cases $|S|$ is based on the two main steps of the checklist generation approach:

$$|S| = |S_1| + |S_2|$$

where $S_1$ is a subset of test cases that affects individual components; and $S_2$ is a subset of test cases that affects data flows between pairs of individual components, which must communicate.

TABLE I
PARTITIONED LIST OF THREATS OF IT-GRUNDSCHUTZ [27]

| THREATS DESCRIPTION | SETS OF TREATS | | | | | | | | | |
|---|---|---|---|---|---|---|---|---|---|---|
| | $Th_{01}$ | $Th_{11}$ | $Th_{21}$ | $Th_{31}$ | $Th_{51}$ | $Th_{02}$ | $Th_{12}$ | $Th_{22}$ | $Th_{32}$ | $Th_{52}$ |
| T 0.01 Fire | x | | | | | | | | | |
| T 0.02 Unfavorable climatic conditions | x | | | | | | | | | |
| T 0.03 Water | x | | | | | | | | | |
| T 0.04 Pollution, dust, corrosion | x | | | | | | | | | |
| T 0.05 Natural disasters | x | | | | | | | | | |
| T 0.06 Environmental disasters | x | | | | | x | | | | |
| T 0.07 Major events in the environment | x | | | | | x | | | | |
| T 0.08 Failure or disruption of the power supply | x | | | | | | | | | |
| T 0.09 Failure or disruption of communication networks | | x | | | | | x | | | |
| T 0.10 Failure or disruption of mains supply | x | | | | | | | | | |
| T 0.11 Failure or disruption of service providers | | | | | | | | x | | |
| T 0.12 Interfering radiation | x | | | | | x | | | | |
| T 0.13 Intercepting compromising emissions | x | | | | | x | | | | |
| T 0.14 Interception of information / espionage | | | | x | | | | | x | |
| T 0.15 Eavesdropping | | x | | | | | x | | | |
| T 0.16 Theft of devices, storage media and documents | x | | | | | | | | | |
| T 0.17 Loss of devices, storage media and documents | | | | | x | | | | | |
| T 0.18 Bad planning or lack of adaptation | | | | | x | | | | | x |
| T 0.19 Disclosure of sensitive information | | | | x | | | | | x | |
| T 0.20 Information or Product from an unreliable source | | | | | x | | | | | |
| T 0.21 Manipulation of hardware and software | | | | | x | | | | | |
| T 0.22 Manipulation of information | | | | x | | | | | | |
| T 0.23 Unauthorized access to IT systems | | | x | x | | | | | | |
| T 0.24 Destruction of devices or storage media | | | | | x | | | | | |
| T 0.25 Failure of devices or systems | | x | | | | | | | | |
| T 0.26 Malfunction of devices or systems | | x | | | | | | | | |
| T 0.27 Lack of resources | x | | | | | x | | | | |
| T 0.28 Software vulnerabilities or errors | | | x | x | | | | | | |
| T 0.29 Violation of laws or regulations | | | | | x | | | | | x |
| T 0.30 Unauthorized use or administration of devices and systems | | x | x | x | | | x | x | x | |
| T 0.31 Incorrect use or administration of devices and systems | | | x | x | | | | x | x | |
| T 0.32 Abuse of authorizations | | | x | x | | | | x | x | |
| T 0.33 Absence of personal | | | | | x | | | | | |
| T 0.34 Attack | x | | | | | | | | | |
| T 0.35 Coercion, extortion or corruption | | | | | x | | | | | |
| T 0.36 Identity theft | | | | | x | | | | | |
| T 0.37 Repudiation of actions | | | | | x | | | | | |
| T 0.38 Abuse of personal data | | | | | x | | | | | |
| T 0.39 Malicious software | | | | x | | | | | | |
| T 0.40 Denial of service | | | | x | | | | | | |
| T 0.41 Sabotage | | | | | x | | | | | |
| T 0.42 Social Engineering | | | | | x | | | | | |
| T 0.43 Replay of messages | | | | x | | | | | | |
| T 0.44 Unauthorized entry to premises | x | | | | | | | | | |
| T 0.45 Data loss | | | | x | | | | | | |
| T 0.46 Loss of integrity of sensitive information | | | | x | | | | | | |

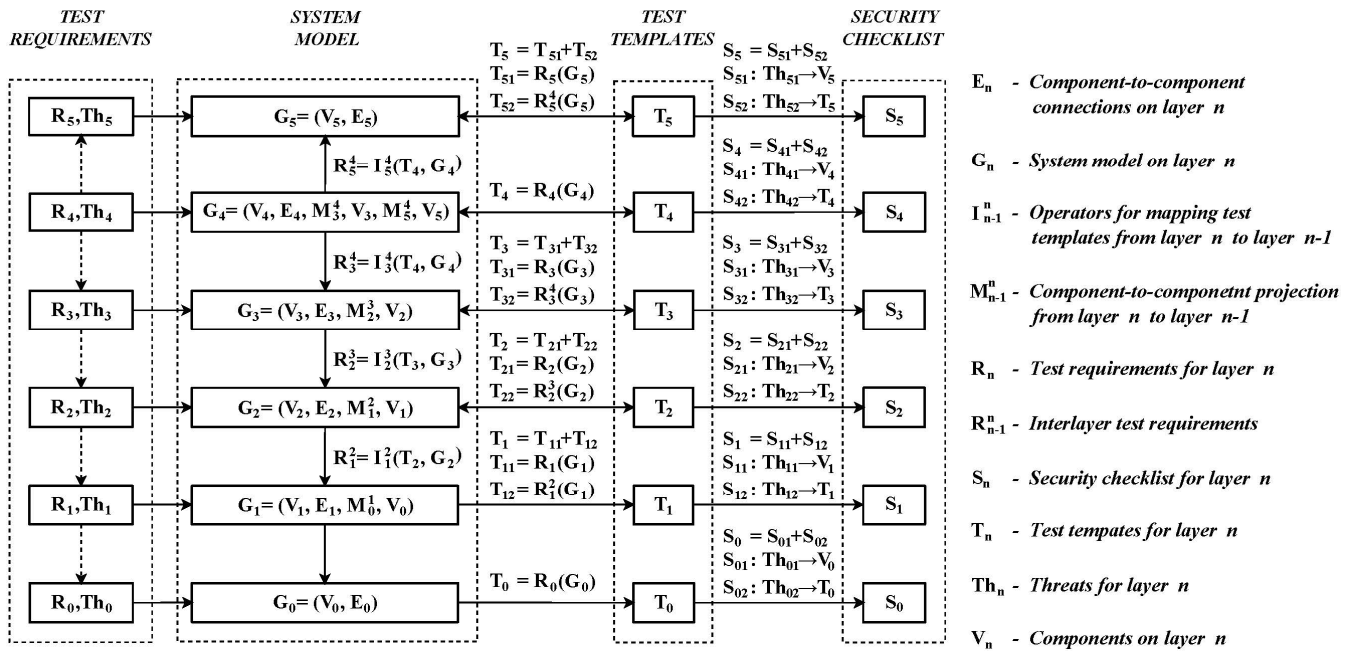

Fig. 3. Graphical representation of the security checklist generation process.

*1)* In the case of individual components, the result is a quite trivial:

$$|S_1| \leq \sum_{n=1}^{N} |Th_{n1}| \, |V_n|$$

*2)* In the case of data flows, there are two possible options:

*2.1) Simple communication systems.* In this case, there is only one possible route between each pair of individual components, which must communicate. For such systems:

$$|T_n| \leq \frac{|V_n|(|V_n| - 1)}{2}$$

And:

$$|S_2| \leq \sum_{n=1}^{N} \left[ \frac{|Th_{n2}||V_n|(|V_n| - 1)}{2} \right]$$

*2.2) Complex communication systems.* For these systems there are some independent routes between each pair of individual components, which must communicate. In this case:

$$|T_n| \leq \alpha \frac{|V_n|(|V_n| - 1)}{2}$$

where $\alpha$ is the number of possible independent routes. In the real engineering word under financial constraints commercial systems are usually based on redundant architecture [26]: $\alpha = 2$. We need to state here – specific areas like the military, nuclear or aerospace industries are beyond the scope this work. Then:

$$|S_2| \leq \sum_{n=1}^{N} |Th_{n2}||V_n|(|V_n| - 1)$$

Table 2 shows the cardinality of the set of threats based on IT-Grundschutz risk management methodology [27].

| Threats subset | Model layer (n) | | | | | |
|---|---|---|---|---|---|---|
| | *0* | *1* | *2* | *3* | *4* | *5* |
| $|Th_{n1}|$ | 15 | 5 | 5 | 13 | - | 13 |
| $|Th_{n2}|$ | 5 | 3 | 4 | 5 | - | 2 |

## IV. CASE STUDY

As a practical example, we have a very simple distributed system (see Fig. 4 – Fig. 7). The service architectural layer and the logical architectural layer are represented in [19] (they are not shown here because of the lack of space).

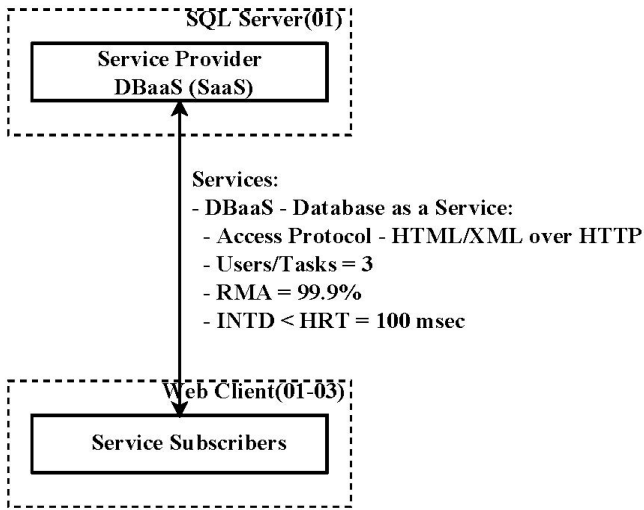

Fig. 4 Simple example of multi-layered system model - functional architecture layer [19]

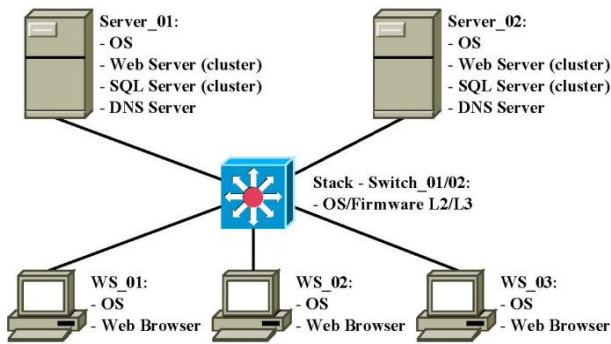

Fig. 5 Simple example of multi-layered system model - physical architecture layer [19]

The security checklist generation process application to the example shows a surprisingly large number of tests required to fully cover even this very simple system – see Table 3.

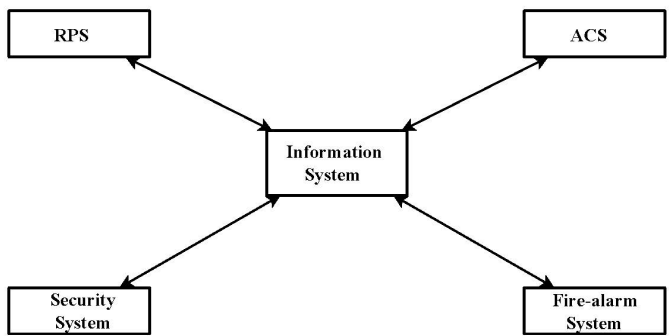

Fig. 6 Simple example of multi-layered system model - engineering environment architecture layer

By and large, this result can easily explain the existence of the huge number of vulnerabilities in commercial systems. Increasing system complexity and fierce market competition on time-to-market and cost make complex security testing of distributed systems really difficult (or even impossible).

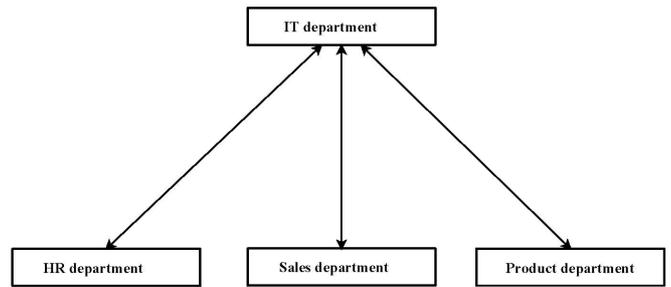

Fig. 7 Simple example of multi-layered system model - social environment architecture layer

TABLE III
APPLICATION OF THE SECURITY CHECKLIST GENERATION PROCESS

| Architectural layers | Security checklists | | | | | |
|---|---|---|---|---|---|---|
| | n | $S_{n1}$ | | $S_{n2}$ | | $|S_n|$ |
| | | $|V_n|$ | $|Th_{n1}|$ | $|T_n|$ | $|Th_{n2}|$ | |
| Social environment | 5 | 4 | 13 | 3 | 2 | 58 |
| Functional | 4 | 2 | - | 1 | - | - |
| System | 3 | 14 | 13 | 15 | 5 | 257 |
| Logical | 2 | 6 | 5 | 7 | 4 | 58 |
| Physical | 1 | 7 | 5 | 6 | 3 | 53 |
| Engineering environment | 0 | 4 | 15 | 4 | 5 | 80 |
| | | | | | Total: | 506 |

## V. CONCLUSIONS

When talking about the security testing of distributed systems, we face the great challenge of the dual nature of these systems. Today every computing and/or communication (network) component is a computer with special operating software. So, it is necessary to pay respect to their software nature and we have to talk about "software-based systems" instead of simply "systems". On the other hand, the distribution character of these systems forces us to consider their network nature. So, test applications should be very flexible to cover distributed systems appropriately and test multiple different aspects with a variety of requirements [28].

Historically, the software part is "a private area" for system and software engineers. Respectively, the network part is "a private area" for network engineers and partially for system engineers. As a consequence, system, software and network engineers have few common models or approaches and even their vocabulary is different [1]. The situation worsens when we talk about industrial control systems. Objectively, they have the same physical nature as every distributed system. But practically, "the private area" of industrial engineers was closed for system and/or network engineers for many years [29].

In this work we determined: (1) a set of typical threats based on standards and actual practices; and (2) a formal model for test generation mission on the basis of the concept of layered networks. The model is a six-layered graph, derived from the system technical specifications, which covers software-based and network-based aspects of distributed systems and their environments – external engineering systems and organization infrastructures (see Fig. 2).

Applying of the requirements-coverage test strategy [19] to the system model provides a set of system objects that need to be protected:

- individual components;
- individual components interaction from the end-user requirements on all architectural layers.

Next, we partitioned the set of typical threats with regard to the model layered structure (see Table 1). The result of the layer-by-layer mapping from this set of threats to the set of protected objects (individual components and their interactions) is the necessary security checklist.

### ACKNOWLEDGMENT

This research has been performed within the scientific activities at the Department of Telecommunication Engineering of the Czech Technical University in Prague, Faculty of Electrical Engineering.